\documentclass{iau}
\usepackage{graphicx}
\usepackage{natbib}

\newcommand{\apj}{ApJ}
\newcommand{\aap}{A\&A}
\newcommand{\mnras}{MNRAS}
\newcommand{\nat}{Nature}
\newcommand{\apjl}{ApJL}
\newcommand{\apjs}{ApJS}

\newcommand{\aapr}{AARev}

\title[Planets spinning up their host stars] 
{Planets spinning up their host stars: a twist on the age-activity relationship}

\author[K.\ Poppenhaeger \& S.\ J.\ Wolk]   
{K.\ Poppenhaeger$^1$
 \and S.\ J.\ Wolk$^1$}

\affiliation{$^1$Harvard-Smithsonian Center for Astrophysics, \\ 
60 Garden Street, \\
02138 Cambridge, 02138 MA, USA \\
email: {kpoppenhaeger@cfa.harvard.edu}}

\pubyear{2013}
\volume{xxx}  
\pagerange{xxx--xxx}
\jname{IAUS 302 - Magnetic fields throughout stellar evolution}
\begin{document}

\maketitle

\begin{abstract}
It is a long-standing question in exoplanet research if Hot Jupiters can influence the magnetic activity of their host stars. While cool stars usually spin down with age and become inactive, an input of angular momentum through tidal interaction, as seen for example in close binaries, can preserve high activity levels over time. This may also be the case for cool stars hosting a Hot Jupiter. However, selection effects from planet detection methods often dominate the activity levels seen in samples of exoplanet host stars, and planet-induced, systematically enhanced stellar activity has not been detected unambiguously so far. We have developed an approach to identify planet-induced stellar spin-up avoiding the selection biases from planet detection, by using visual proper motion binaries in which only one of the stars possesses a Hot Jupiter. This approach immediately rids one of the ambiguities of detection biases: with two co-eval stars, the second star acts as a negative control. We present results from our ongoing observational campaign at X-ray wavelengths and in the optical, and present several outstanding systems which display significant age/activity discrepancies presumably caused by their Hot Jupiters.

\end{abstract}

\firstsection 
\section{Introduction}

Almost all planet-hosting stars known today are cool stars (spectral types F-M). All cool stars display stellar activity -- a summarizing term for the occurrance of magnetic phenomena including flares, spots, and coronal high-energy emission. The magnetic activity of planet-hosting stars is an important factor to understand the evolution of exo\-pla\-nets. High-energy irradiation of close-in planets can lead to atmospheric evaporation \citep{Vidal-Madjar2003, Lecavelier2010}, and coronal mass ejections and the stellar wind can strip away parts of the planetary atmosphere \citep{Penz2008}. The stability and chemistry of exoplanetary atmospheres are both influenced by these stellar activity phenomena. Recent transit observations in the UV \citep{Bourrier2013} and in X-rays \citep{Poppenhaeger2013} have demonstrated that atmospheres of Hot Jupiters are strongly extended. Modelling of such atmospheres suggests that time-variable phenomena like bow-shocks and variations in the stellar wind may play a role, too \citep{Vidotto2012}.

Stellar activity levels of stars with and without planets have received much scrutiny in recent years. Activity is well-known to be a function of stellar rotation and therefore, due to magnetic breaking, of stellar age. Stellar X-ray luminosity therefore declines with stellar age, as shown in Fig.~\ref{age-activity}, left panel.\footnote{For a summary of age-activity relations using the fractional X-ray luminosity $L_X/L_{bol}$ instead of $L_X$, see \cite{Jackson2012}.} Several studies have noted that exoplanets may also have an influence on the stellar magnetic activity. The underlying idea is that massive close-in exoplanets should be able to interact with their host star magnetically and/or tidally \citep{Cuntz2000, Shkolnik2005, Kashyap2008}. However, because stellar activity effectively masks signals used to detect exoplanets, one has to be careful not to interpret intrinsic biases in samples of planet-hosting stars as a physical interaction effect \citep{Poppenhaeger2010, Poppenhaeger2011, Miller2012}. 

A useful way to test for activity enhancements caused by Hot Jupiters is to look at wide stellar binaries, in which only one star hosts a planet massive and close enough for strong tidal interaction. The star without a planet then acts as a negative control, because its activity evolution should be unperturbed by any planetary effects. One can thus test if both stars display magnetic activity levels appropriate for their common stellar age (taking into account slight differences due to spectral type). We are conducting an observational program using X-ray data and optical spectra to test for such activity differences.

\section{Initial results}

\begin{figure}[t!]
\begin{center}
\begin{minipage}[b]{0.45\textwidth}
\includegraphics[width=1.0\textwidth]{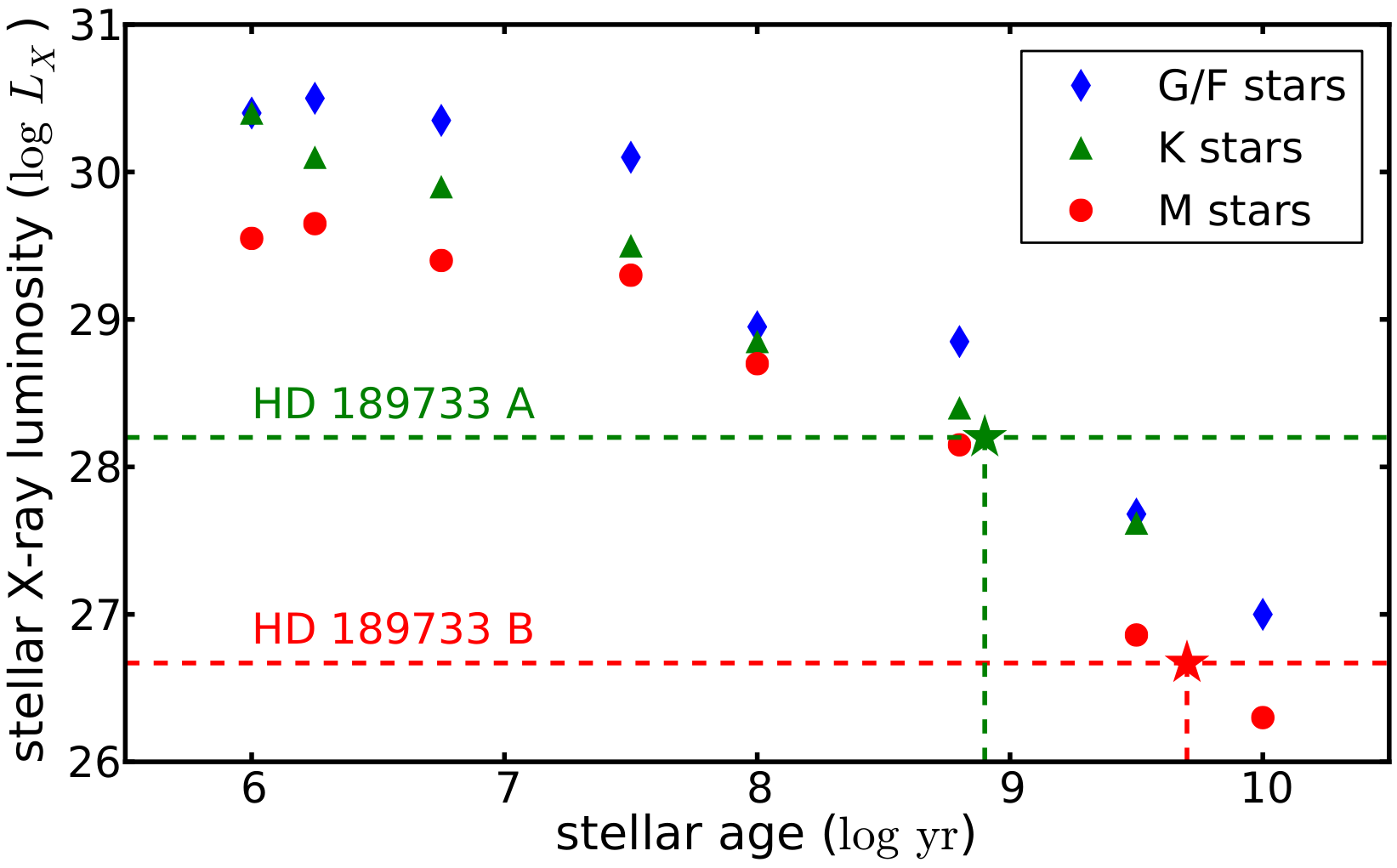}

\vspace*{0.9cm}
\end{minipage}
\hspace{0.1cm}
\begin{minipage}[b]{0.5\textwidth}
\includegraphics[width=1.0\textwidth]{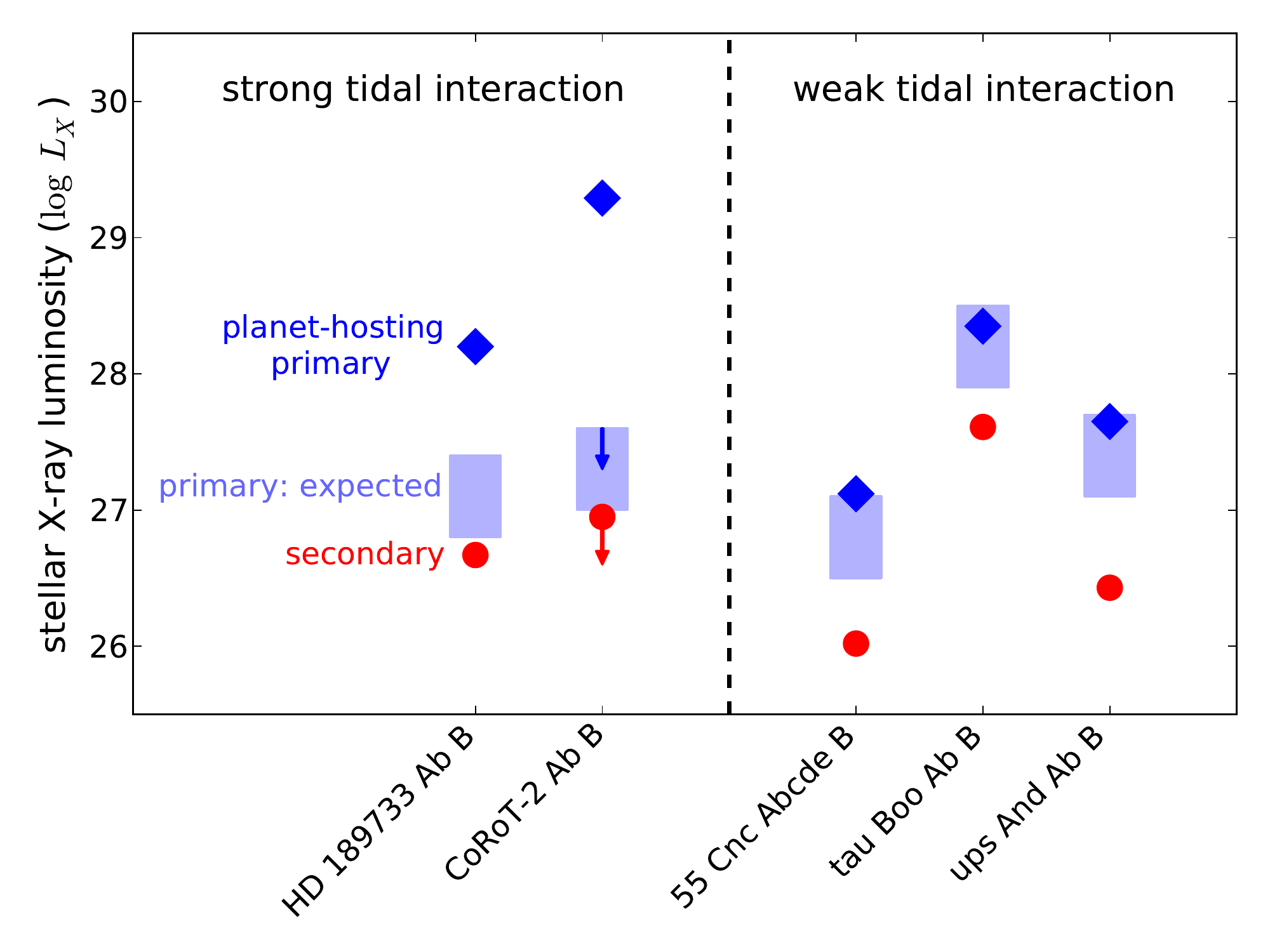}
\end{minipage}
\end{center}

\caption{{\bf Left:} Magnetic breaking causes cool stars to spin down over time, which can be measured as a decline in magnetic activity indicators such as X-ray emission.$^{*}$ {\bf Right:} In wide stellar binaries in which one of the stars hosts an exoplanet, some of the planet-hosting stars display a much higher magnetic activity level than expected for their age. This is the case for systems with strong tidal interaction between planet and host star.
\medskip
\newline
$^{*}$\,Age-activity data from \cite{Preibisch2005}, \cite{Jeffries2006}, \cite{LiefkeSchmitt2004}, \cite{Damiani2004}, \cite{Engle2011}, \cite{Hawley1994}. }
\label{age-activity}
\end{figure}

\begin{table}[ht!]
\begin{tabular}{l c c c c c c c c}
\hline\hline
&&&&&&&&\vspace*{-0.2cm}\\

System	& SpT A	& SpT B	& $a_{sem}$	& $M_P$		& $\log L_{X,\,B}$	& age$_B$	& $\log L_{X}^{A\,(exp.)}$	& $\log L_{X}^{A\,(obs.)}$ \\
	& 	& 	& (AU)		& ($M_{Jup}$)	& (erg/s)	& (Gyr)		& (erg/s)			& (erg/s)	\\
\hline
HD\,189733\,Ab\,B		& K1V	& M4V	& 0.03	& 1.138		& 26.67		& 5	& 27.1	& 28.2	\\
CoRoT-2\,Ab\,B			& G9V	& K9V	& 0.01	& 2.0		& $<$26.95	& 3	& 27.3	& 29.3	\\
55\,Cnc\,Abcde\,B		& K2V	& M3V	& 0.03	& $>$0.0001	& 26.02		& 10	& 26.8	& 27.12	\\
$\tau$\,Boo\,Ab\,B		& F6V	& M3V	& 0.03	& $>$1.0	& 27.61		& 2	& 28.2	& 28.35	\\
$\upsilon$\,And\,Ab\,B		& F7V	& M4V	& 0.03	& $>$1.0	& 26.43		& 7	& 27.2	& 27.65	\\
\hline

\end{tabular}
\caption{Observed X-ray luminosities of the planet-hosting primaries and the secondaries which are not known to possess planets. The X-ray luminosity of the secondary is used to estimate the age of the system. Assuming that both stelar components are co-eval, we can derive an {\it a priori} estimate how X-ray luminous the primary should be (column $\log L_{X,\,A\,(exp.)}$). The actual observed X-ray luminoities of the primaries are given in column $\log L_{X,\,A\,(obs.)}$. }
\label{Lxtable}
\end{table}

We have collected observational data for five exoplanet systems with wide stellar binaries so far, using {\it XMM-Newton} and {\it Chandra} for X-ray observations of the stellar corona, as well as FLWO's TRES spectrograph for optical high-resolution spectra. We list the observed X-ray luminosities for both stellar components of each system in Table~\ref{Lxtable}. We estmate the age of the system from the X-ray brightness of the secondary (non-planet-hostimg) star. This yields an estimate for the X-ray luminosity of the primary (planet-hosting) star, assuming that both stars are co-eval. More detailed analyses of the spectral X-ray properties of some of the stars can be found in \cite{Pillitteri2010} and \cite{Poppenhaeger2013} for HD~189733 A and B, \cite{Schroeter2011} for CoRoT-2 A, \cite{Poppenhaeger2012AN} for $\tau$ Boo and \cite{Poppenhaeger2011} for $\upsilon$ And A. 

For three of the systems we find that the expected and observed X-ray luminosities for the planet-hosting stars are in agreement; for two systems, however, the expected and observed X-ray luminosities disagree by an order of magnitude or more; see Fig.~\ref{age-activity}, right panel. These systems both host a Hot Jupiter in a very close orbit, and the host star is a late G or K dwarf. For the systems in which no discrepancy is found, the innermost planet is either small (55 Cnc) or the host star is an F star with a very thin outer convective envelope ($\tau$~Boo and $\upsilon$~And).

We interpret our initial findings as a manifestation of tidal interaction between the planet and the host star. Theoretical studies show that Hot Jupiters induce tidal bulges on the host star; cool stars can dissipate the energy contained in the bulges much more effectively than hot stars due to turbulent eddies in the convective envelopes, see \cite{Zahn2008} and \cite{Torres2010}. Observationally, stars with substantial convective envelopes have also been found to have exoplanets in orbits with low orbital obliquity, which is also believed to be a consequence of tidal interaction \citep{Winn2010, Albrecht2012}. 

The Hot Jupiters in the HD~189733A and CoRoT-2A systems may therefore have inhibited the spin-down of their host stars, by transferring angular momentum from the planetary orbit into the stellar spin. 

To test this hypothesis further, we are collecting high-resolution optical spectra for our sample to measure the projected rotational velocities $v\sin i$ of the stars as well as chromospheric activity indicators such as Ca II H and K and H $\alpha$.

\section{Conclusion}

Our initial data suggests that there is indeed a large difference in activity levels for systems in which the planet exerts a strong tidal interaction on its host star, while systems with weak tidal interaction do not show elevated activity levels. Observations of a larger number of systems which can be tested for possible tidal effects from the planet are under way.

The consequences of planet-induced stellar high activity are substantial for understanding exoplanetary systems and, on a larger scale, stellar evolution. For systems with Hot Jupiters, age estimates can no longer be derived from stellar rotation and activity; also, the time-integrated high-energy irradiation of exoplanets will be much higher due to the prolonged high activity of the host stars, so that planetary evaporation may have had a much stronger influence on the planet zoo we observe today than current models suggest. In terms of stellar evolution, stellar population synthesis models are just beginning to include the star's spin as a parameter. First results show that the stellar rotation is a fundamental parameter for its evolution \citep{Levesque2012}, so that it is crucial to know if Hot Jupiters significantly change the spin of their host stars, which comprise ca.\ 1\% of all cool stars \citep{WrightMarcy2012}.


\end{document}